# Performance Analysis of Wavelet Based MC-CDMA System with Implementation of Various Antenna Diversity Schemes

Md. Matiqul Islam, M. Hasnat Kabir and Sk. Enayet Ullah

Department of Information and Communication Engineering, University of Rajshahi
Rajshahi-6205, Bangladesh

*Abstract:* The impact of using wavelet based technique on the performance of a MC-CDMA wireless communication system has been investigated. The system under proposed study incorporates Walsh Hadamard codes to discriminate the message signal for individual user. A computer program written in Mathlab source code is developed and this simulation study is made with implementation of various antenna diversity schemes and fading (Rayleigh and Rician) channel. Computer simulation results demonstrate that the proposed wavelet based MC-CDMA system outperforms in Alamouti (two transmit antenna and one receive antenna) under AWGN and Rician channel.

## 1. Introduction

In 4$^{th}$ generation wireless mobile communication it is the challenge to provide high-data-rate up to 100 Mb in a spectrally efficient manner with utilizing the available limited bandwidth for high quality communication. Multicarrier code division multiple access (MC-CDMA)[1], [2], which is the close union of OFDM and CDMA, has drawn a lot of advantage for its robustness to channel dispersion. However, due to the limitation of OFDM and CDMA, a great attention has been focus on MC-CDMA. This technique is capable to accommodate a higher number of users as compared to CDMA. Three types of multicarrier CDMA schemes are available in literature namely: 1) MC-CDMA, 2) MC direct-sequence CDMA, and 3) Multitone CDMA [3]. MC-CDMA signal can be produced by multiplexing some orthogonal subcarriers as well as OFDM. As a matter of fact, DFT based conventional MC-CDMA does not concentrate energy effectively [4]. Therefore, the performance of synchroning error is reduced due to the inter-channel –interference (ICI) and inter-symbol interference (ISI). As a result, an efficient MC-CDMA system is required to achieve better performance. Several improvements of this system have been proposed. Among them, wavelet based MC-CDMA [5], [6] shows much attraction to the researcher for stronger ability to combat multipath interference (MPI) and ISI than conventional one. It has the advantage of avoiding the influence of delayed waves and eliminating the intersymbol interference (ISI) using a guard interval in multipath environment. Nevertheless, interference can be removed by ensuring smaller sub channel bandwidth than the channel coherence bandwidth (CB).
Recently wavelet has been developed as a new signal processing tool which enables the analysis on several timescales of the local properties of compiles signals. An improvement of wavelet based multicarrier code division multiple access is proposed by Salih M. et al [7]. On the other hand, Saad N. [8] has introduced Discrete Multiwavelet Transform (DMWT), a new structure of MC-CDMA. However, there is still some opportunities to improve the performance of wavelet based MC-CDMA system. Salih M. [9] has suggested that the performance can be improved by combining MC-CDMA with antenna arrays at the transmitter and the receiver to increase the diversity gain, resulting in a Multiple-input Maltiple-output (MIMO) configuration. So, it is meaningful to implement antenna diversity scheme for measuring the performance of wavelet based MC-CDMA system.

Our research paper deals with multi-carrier transmission scenario using wavelet transform instead of OFDM to work in a fading channel environment. Alamouti STBC is used as antenna diversity scheme. It is widely used for mitigating the effect of multipath fading. This scheme will be described in detail latter. The results are compared with no diversity as well as maximal radio combiner (MRC).

The paper organizes as follows: section 2 describes the Alamouti STBC scheme which is used in this research. The proposed model is described in section 3 as system description. In section 4, the simulation results of the proposed wavelet based MC-CDMA system are presented.

## 2. Alamouti STBC Scheme

In 1998, Alamouti has presented a simple two-branch transmit diversity scheme. Using two transmit antennas and one receive antenna the scheme provides the same diversity order as maximal-ratio receiver combining (MRRC) with one transmit antenna and two receive antennas [10], [11]. Figure 1 shows the block diagram of base band representation of the Alamouti's two branch transmit diversity scheme. The scheme is defined by the three functions such as encoding and transmission sequence of information symbols at the transmitter, combining scheme at the receiver and decision rule for maximum likelihood detection

In encoding and transmission sequence, two signals are simultaneously transmitted from the two antennas at a given symbol period. The signal transmitted from antenna zero is denoted by $s_o$ and from antenna one by $s_1$. During the next symbol period, signal ($-s_1^*$) is transmitted from antenna zero and signal $s_o^*$ is transmitted from antenna one where * is the complex conjugate operation. In Alamouti scheme, the encoding is done in space and time (space–time coding) and such encoding may also be done in space and frequency.

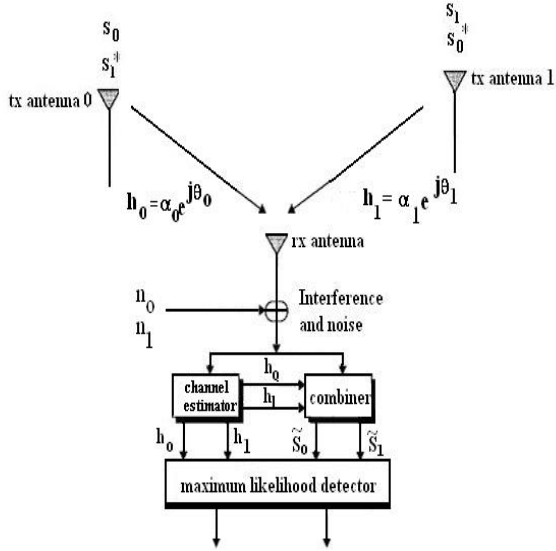

**Figure 1.** Conceptual block diagram of Alamouti space-time block coding scheme.

The channel at time t may be modeled by a complex multiplicative distortion $h_o(t)$ for transmit antenna zero and $h_1(t)$ for transmit antenna one. Assuming that fading is constant across two consecutive symbols, we can write

$$h_o(t) = h_o(t+T) = h_o = \alpha_o e^{j\theta_o}$$
$$h_1(t) = h_1(t+T) = h_1 = \alpha_1 e^{j\theta_1} \quad (3)$$

Where, T is the symbol duration. The received signals can then be expressed as

$$r_o = r(t) = h_o s_o + h_1 s_1 + n_o$$
$$r_1 = r(t+T) = -h_o s_1^* + h_1 s_0^* + n_1 \quad (4)$$

Where $r_o$ and $r_1$ are the received signals at time t and t+ T and are $n_o$ and $n_1$ complex random variables representing receiver noise and interference. The combiner shown in Figure 1 builds the following two combined signals that are sent to the maximum likelihood detector:

$$\tilde{s}_o = h_o^* r_o + h_1 r_1^*$$
$$\tilde{s}_1 = h_1^* r_o - h_o r_1^* \quad (5)$$

Substituting (1) and (2) into (3), we get

$$\tilde{s}_o = (\alpha_o^2 + \alpha_1^2)s_o + h_o^* n_o + h_1 n_1^*$$
$$\tilde{s}_1 = (\alpha_o^2 + \alpha_1^2)s_1 - h_o n_1^* + h_1^* n_o \quad (6)$$

These combined signals are then sent to the Maximum likelihood detector which, for each of the signals $s_o$ and $s_1$ uses the decision rule expressed in (7) or (9) for PSK signals Choosing $s_i$ if

$$(\alpha_o^2 + \alpha_1^2 - 1)|s_i|^2 + d^2(\tilde{s}_o, s_i) \backslash$$
$$\leq (\alpha_o^2 + \alpha_1^2 - 1)|s_k|^2 + d^2(\tilde{s}_o, s_k), \forall i \neq k \quad (7)$$

Where, $d^2(\tilde{s}_o, s_k)$ is the squared Euclidean distance between signals $\tilde{s}_o$ and $s_k$ calculated by the following expression

$$d^2(\tilde{s}_o, s_k) = (\tilde{s}_o - s_k)(\tilde{s}_o^* - s_k^*) \quad (8)$$

Choosing $s_i$

$$d^2(\tilde{s}_o, s_i) \leq d^2(\tilde{s}_o, s_k), \forall i \neq k \quad (9)$$

## 3. System Description

The wavelet based CDMA transmitter includes M-branches and each of them are consist with an up-sampler followed by a synthesis filter. The impulse response of this filter is derived from the wavelet orthogonal condition which generates a specific wavelet pulse. We have used conventional wavelet based MC-CDMA system as describe in literature [6] in addition to antenna diversity.

The proposed simulated wavelet based MC-CDMA system with implementation of various antenna diversity schemes is depicted in Figure 2. The synthetically generated binary bit stream for four users are fed into time division multiplexer (parallel to serial converter).The serially converted multiplexed binary data are converted into complex symbols using BPSK digital modulation and these symbols are converted into serial to parallel form and feed into copier section where these symbols are multiplied with assigned individual Walsh Hadamard code. The coded symbols are passed through inverse wavelet transformation and eventually fed into Alamouti space time block encoder. Its output provides complex symbols in two diversity channels.

In receiving section, the two transmitting message signals are processed with perfect knowledge of channel information in diversity combiner. There are M branches in the receiver and each of them is consist with a filter followed by a down-sampler. The inverse operations are performed to recover the original signal.

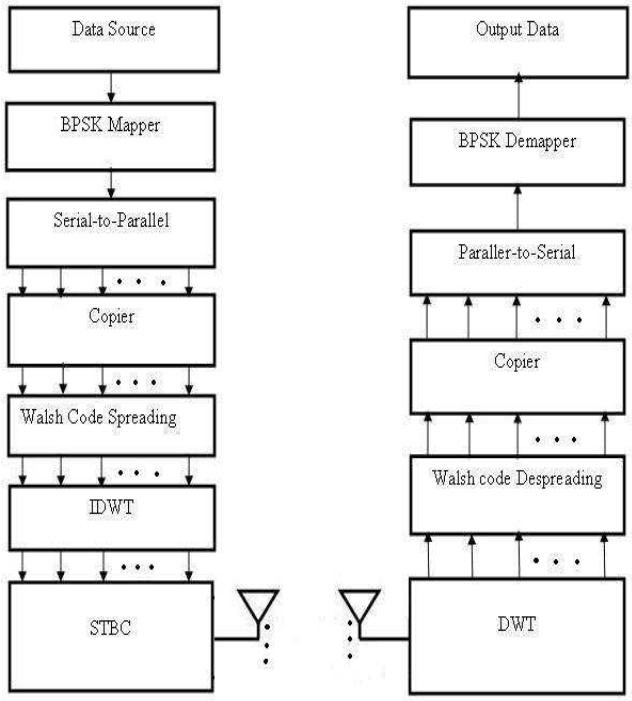

**Figure 2.** Wavelet based MC-CDMA system with implementation of various antenna diversity schemes.

## 4. Results and Discussion

During simulation, the perfect synchronization between the transmitted and the received signals has been considered with available channel state information (CSI) at the receiver. The channel coefficients are assumed to be static during two consecutive transmitted symbols in the diversified channels. The simulation parameters are shown in table 1. The performance of the proposed system was tested in different channels. Figure 3 shows the BER performance of the wireless communication system with and without implementation of diversity scheme under BPSK digital modulation techniques in AWGN channel for wide range of SNR from 0 dB to 20 dB.

Table 1. The parameters of simulation model.

| Parameter | Parameter Value |
|---|---|
| No. of bits used for synthetic data | 1,0 |
| No. of user | 4 |
| SNR | 0-20 db |
| Spreading code | Walsh-Hadamard Code, Length 4 |
| Wireless channel | AWGN, Rayleigh fading, Rician channel |
| Maximum Doppler shift in Rayleigh fading channel | 130 dB |
| Maximum Doppler shift in Rician fading channel | 100dB |

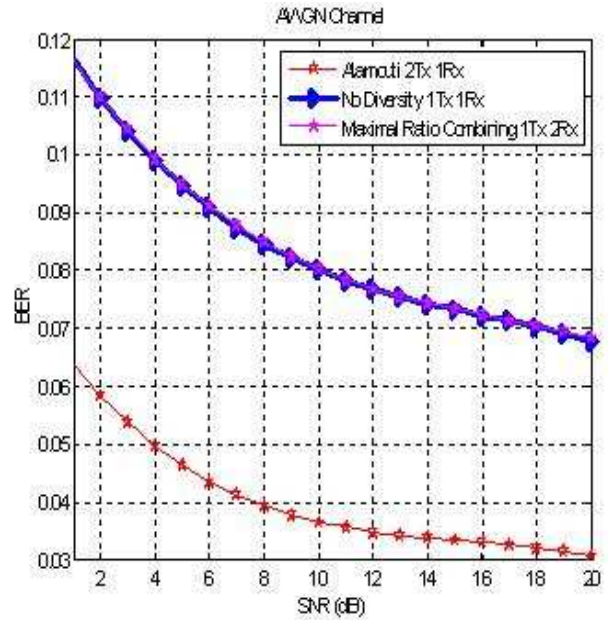

**Figure 3.** Wavelet based MC-CDMA system with implementation of various antenna diversity schemes in AWGN channel.

The system performance is achieved significantly with deployment of Alamouti diversity scheme as compared to no diversity and MRC scheme. In a typically assumed SNR value of 5 dB with BPSK modulation, it is observed that the bit error rates are 0.0463 and 0.0950 for Alamouti diversity and no diversity (direct transmission) schemes, respectively. This means that a gain of 3.12 dB is obtained by the proposed model. However, MRC shows similar nature as about direct transmission under AWGN.

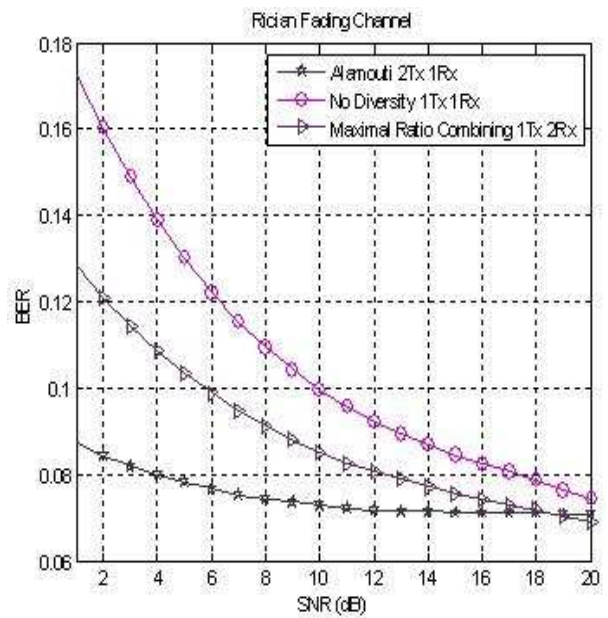

**Figure 4.** Wavelet based MC-CDMA system with implementation of various antenna diversity schemes in Rician fading channel.

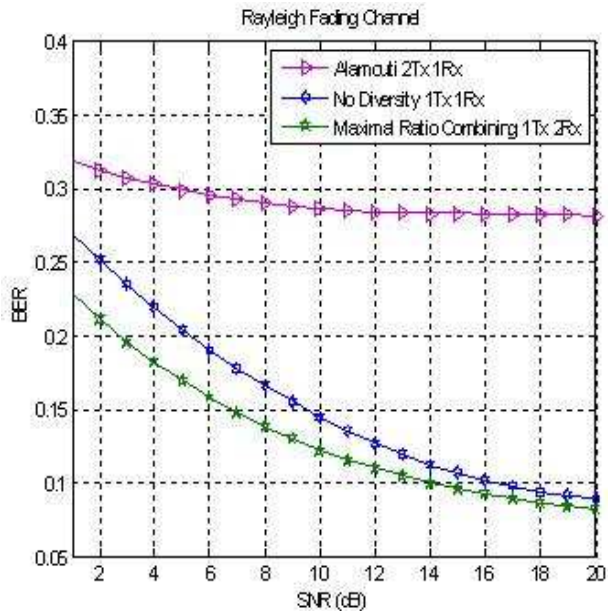

**Figure 5.** Wavelet based MC-CDMA system with implementation of various antenna diversity schemes in Rayleigh fading channel.

Simulations are also carried out to investigate the system performance under different channel. The result shown in figure 4 is the BER performance of the proposed communication system under in Rician fading channel. It is seen from this figure that the BER values at SNR = 6 dB are approximately 0.0767, 0.0987 and 0.1222 for Alamouti, MRC and no diversity, respectively. It can be pointed out from here that under multipath fading channel, Alamouti shows better BER performance than others. Two time better performance can be achieved using Alamouti as compared to no diversity. This is because the effect of interference caused by multipath fading is mitigated and interference due to ICI and ISI are strongly suppressed. On the other hand, figure 5 shows the BER performance with similar modulation technique in Rayleigh fading channel. Here Maximal Ration Combining (MRC) shows better performance as compared to others. It can be explained that the diversity order increases. Here the effective channel is concatenating the information from 2 receives antennas and as we know that m receive antennas, the diversity order for 2 transmit antenna Alamouti STBC is 2m i.e. the diversity order is 4.

## 5. Conclusion

A wavelet based MC-CDMA with antenna diversity scheme is proposed in this paper. Alamouti scheme has used as the antennal diversity. The simulation results show that the BER performance is better in proposed system using Alamouti scheme under AWGN and Rician channel where as it is worse under Rayleigh fading channel.